\documentclass[runningheads]{llncs}
\usepackage[T1]{fontenc}

\usepackage{cite}
\usepackage{amsmath,amssymb,amsfonts}
\usepackage{algorithmic}
\usepackage{graphicx}
\usepackage{textcomp}
\usepackage{xcolor}

\def\BibTeX{{\rm B\kern-.05em{\sc i\kern-.025em b}\kern-.08em
    T\kern-.1667em\lower.7ex\hbox{E}\kern-.125emX}}

\begin{document}
\title{Harmonizing Community Science Datasets to Model Highly Pathogenic Avian Influenza (HPAI) in Birds in the Subantarctic}
%
%
\author{Richard Littauer\orcidID{0000-0001-5428-7535} \and
Kris Bubendorfer\orcidID{0000-0003-4315-8337}}
\authorrunning{Littauer and Bubendorfer}
\titlerunning{Modelling Highly Pathogenic Avian Influenza (HPAI) in the Subantarctic}
%
\institute{School of Engineering and Computer Science, Te Herenga Waka Victoria University of Wellington, New Zealand.
\email{{richard.littauer,kris.bubendorfer}@ecs.vuw.ac.nz}\\
}
\maketitle







\begin{abstract} 

Community science observational datasets are useful in epidemiology and ecology for modeling species distributions, but the heterogeneous nature of the data presents significant challenges for standardization, data quality assurance and control, and workflow management. In this paper, we present a data workflow for cleaning and harmonizing multiple community science datasets, which we implement in a case study using eBird, iNaturalist, GBIF, and other datasets to model the impact of highly pathogenic avian influenza in populations of birds in the subantarctic. We predict population sizes for several species where the demographics are not known, and we present novel estimates for potential mortality rates from HPAI for those species, based on a novel aggregated dataset of mortality rates in the subantarctic.


\end{abstract}





\section{Introduction}
\label{sec:intro}

Community science, citizen science, or public participation in science all rely on amateur observations being used to further scientific goals or to provide research data. The widespread adoption and growth of community science has resulted in a flood of accessible, large-scale, crowd-sourced observational datasets -- with projects in physics, life science, medicine, ecology, biology, and many other fields~\cite{Vohland21}.  



Some of these platforms are massive. For instance, eBird 
\cite{sullivan2009ebird, SULLIVAN201431} contains over a billion observational records \cite{ebird1Billion}, with over a million users and 100\,m checklists \cite{ebirdHomepage}. iNaturalist \cite{van2018inaturalist} has almost 250\,m observations, with over 3\,m observers for over 50\,k species \cite{iNaturalist2025}.

While these large datasets offer immense research potential, the inherent heterogeneity of structure, data quality, curation standards and procedures presents significant challenges that limit their impact as research data.  Specifically, community science datasets typically consist of observations of differing quality, from observers of differing skill, using differing protocols~\cite{NEATECLEGG2020108653}. Quality assurance and quality control vary depending on the platform, and the skill, training, and resources of the editors or data reviewers on the platforms. Cleaning processes for using the data are not standardized, and best practices may not cover all environments or use cases. Nevertheless, these datasets as they currently stand are used to produce scientific research and as a consequence to inform public policy globally.

It is possible to mitigate some of these issues by harmonizing data from several community science datasets.  To this end, in this paper, we present and discuss a case study that aggregates and harmonizes data from the two largest community science platforms, eBird and iNaturalist, along with some smaller, relevant datasets to the study area. We utilize this harmonized data to project population sizes for five species of birds -- Black-browed Albatross (\textit{Thalassarche melanophris}), Brown Skua (\textit{Stercorarius antarcticus}), Kelp Gull (\textit{Larus dominicanus}), King Penguin (\textit{Aptenodytes patagonicus}), and the Wandering Albatross complex (\textit{Diomedea} spp.) -- in regions where traditional datasets are limited in their size, frequency, and scope. 

We also combine the aggregated data with a small dataset of mortality statistics for birds infected by Highly Pathogenic Avian Influenza (HPAI or bird flu), which we sourced from recent reports. HPAI is decimating bird species globally, and has only recently arrived in Antarctica and the South Georgia and South Sandwich Islands (SGSSI) in the subantarctic Atlantic Ocean, and at the Prince Edward Islands (PEI) in the subantarctic Indian Ocean. HPAI has yet to reach Oceania, which includes Australia and New Zealand (NZ). By using these statistics, we are able to project possible losses for bird populations in the subantarctic islands of New Zealand (SINZ). Ours are the first known published projections for mortality rates for these islands. 

While there are many studies using community science datasets to show population metrics, there are few studies we could find which use that data to extrapolate population sizes and predict epidemiological patterns. As well, few studies discuss harmonizing data sources in aggregate, using reproducible, versioned methods. By combining datasets in an informed, structured way, we show that similar results can be attained without compromising the underlying data. 
Existing biases cannot be entirely removed, but informed mining of multiple datasets can increase research reliability by mitigating artifacts of individual datasets, leading to more robust outcomes than if one uses a single data source.

\section{Background and Motivation}
\label{sec:background}

\label{sec:community-science}

Community science platforms rely upon experts to review and curate the data provided by volunteers. For eBird \cite{sullivan2009ebird}, the quality control system uses automatic filters to flag unusual observations for review by regional reviewers, who are largely volunteers. The reviewers follow up with individuals to verify their sightings, and may approve entries to be included in the eBird Observational Database (EOD). Reviewers cannot remove observers entries from their personal accounts, but only verified data is entered into the EOD. Researchers may download data, which is updated monthly, from the website. On a yearly basis, this data is also uploaded into the Global Biodiversity Information Facility (GBIF) \cite{gbif}, an aggregator of many datasets that provides the option for researchers to segment, filter, and download datasets. These datasets are also linked to a DOI for reproducibility and citation purposes.

On iNaturalist \cite{van2018inaturalist}, reviewers manage the global taxonomy and unusual observations, but any user can review any observation. If an observation has more than two identifications attached to it, and if a majority of identifications by users are convergent, the data is considered 'research grade' (RG). All data can be downloaded from iNaturalist's website, but only RG observations are also synced with GBIF. iNaturalist demands that observations have either a photo or an audio recording to be considered RG, which acts as a data quality control. eBird observations, on the other hand, do not require evidential proof, although flagged species normally require a comment, photo, or audio to be approved by reviewers.

These systems work at scale, but they are prone to certain biases. For birds, some of these issues are well known \cite{ebirdBestPractices}:
\begin{itemize}
\item \textbf{Sampling bias}. There are more observations where people are, and when people are free to go outside (for instance, on the weekends in many cultures \cite{Courter2013-gg}). Further, some birds are more charismatic -- for instance, endemic birds are logged more than other birds \cite{miskelly2019endemic}.
\item \textbf{Skill bias}. Beginners, amateurs, and experts all use the same tools, and there is no confidence level for reporting \cite{johnston2018estimates}. Even among experts, there is large variation between observers when counting birds -- for instance, up to 47\% variation in numbers for King Penguins \cite{foley2018king}).
\item \textbf{Detection bias}. Some birds are more easily found near humans (House Sparrows, versus albatrosses at sea) \cite{stoudt2022identifying}, are larger and more easily seen \cite{callaghan2021large}, or have different habits and habitats influencing detection (nocturnal owls versus diurnal woodpeckers).
\item \textbf{Duplicated effort}. For eBird, duplicate checklists can sometimes be filtered out. But on both eBird and iNaturalist, if individuals share a bird they saw without sharing metadata with each other, the observation is not filtered out automatically, resulting in skewed distributions. A single rare bird may be reported thousands of times \cite{pease2023steller}.
\item \textbf{Over-reliance on AI}. eBird has released Merlin Bird ID \cite{merlin}, a tool to automatically identify bird sounds which is trained on audio in the EOD. iNaturalist also has an automatic identification algorithm \cite{iNatIDPage} which presents options for identification to every user when they upload a photo. Both of these can present inaccurate identifications, but na\"ive users may rely on the tools and report the wrong species without confirmation.
\item \textbf{Gamification}. Both platforms show lists of users who log the most observations, encouraging healthy competition in order to drive more observations. However, this can result in competitive gaming of the system, and lead to poor data overall~\cite{Anderson09,Sarmenta01}
\end{itemize}

It is generally up to the researcher to control for these biases. Community science platforms allow their data to be downloaded and used, often for free and with permissive data licenses. The platforms may strongly suggest filtering the data before use, such as in the eBird Best Practices Guide \cite{ebirdBestPractices}. Community science is not limited to platforms. For instance, the Royal Naval Birdwatching Society (RNBWS) has forms that its members (Royal Navy personnel) can use to log birds. These are then aggregated into a Microsoft Excel \cite{excel} database. Reviewing the data for quality assurance may happen by the secretary, but is not systematically logged. On the other side of the spectrum, GBIF may have aggregated data of very high quality, such as identifiable museum specimens.

There is nascent field validating statistical models run on community science datasets (c.f. \cite{la2018opportunities, ebirdBestPractices, carlen2024framework, della2024improving}).  Overwhelmingly, these papers urge caution when applying models, as the data can show spurious results, even after cleaning the data. In 2013, one study showed that only 12\% of studies in ecology focus on modeling, while almost none focus on statistics \cite{carmel2013trends}. Community science datasets are fundamentally ecological datasets with an amateur observer bias, and while ecology as a field has shifted in the last decade, there are still unanswered questions on how community science data is applied in practice. 

Here, we outline a workflow for cleaning and aggregating different datasets, based on the following case study.

\section{Case Study: HPAI H5Nx strains in the Southern Oceans}

HPAI has decimated bird populations globally. It has caused unprecedented mass mortality events in South America \cite{leguia2023highly, ulloa2023mass}, North America \cite{avery2024wild}, and Europe \cite{falchieri2022shift, tremlett2024declines}. HPAI has recently spread to the Falkland Islands,
SGSSI \cite{Bennison01102024},
the Crozets and
Kerguelen \cite{clessin2025mass},
PEI (including Marion) \cite{marion2025web}, 
and Antarctica \cite{banyard2024detection}.
As of this writing, it has not yet been recorded in Australia \cite{wildlife2025au}, NZ \cite{mpi2025web}, 
or anywhere else in Oceania. HPAI refers to any bird flu, while the H5Nx strains specifically are the cause of the current virulent wave. Other strands of HPAI may be found in some populations (for instance, the recent H7N6 outbreak in Otago, NZ \cite{mpi2025web}). Here, we refer only to the H5N1 and H5Nx strains.

\begin{figure}
    \centering
    \includegraphics[width=0.6\linewidth]{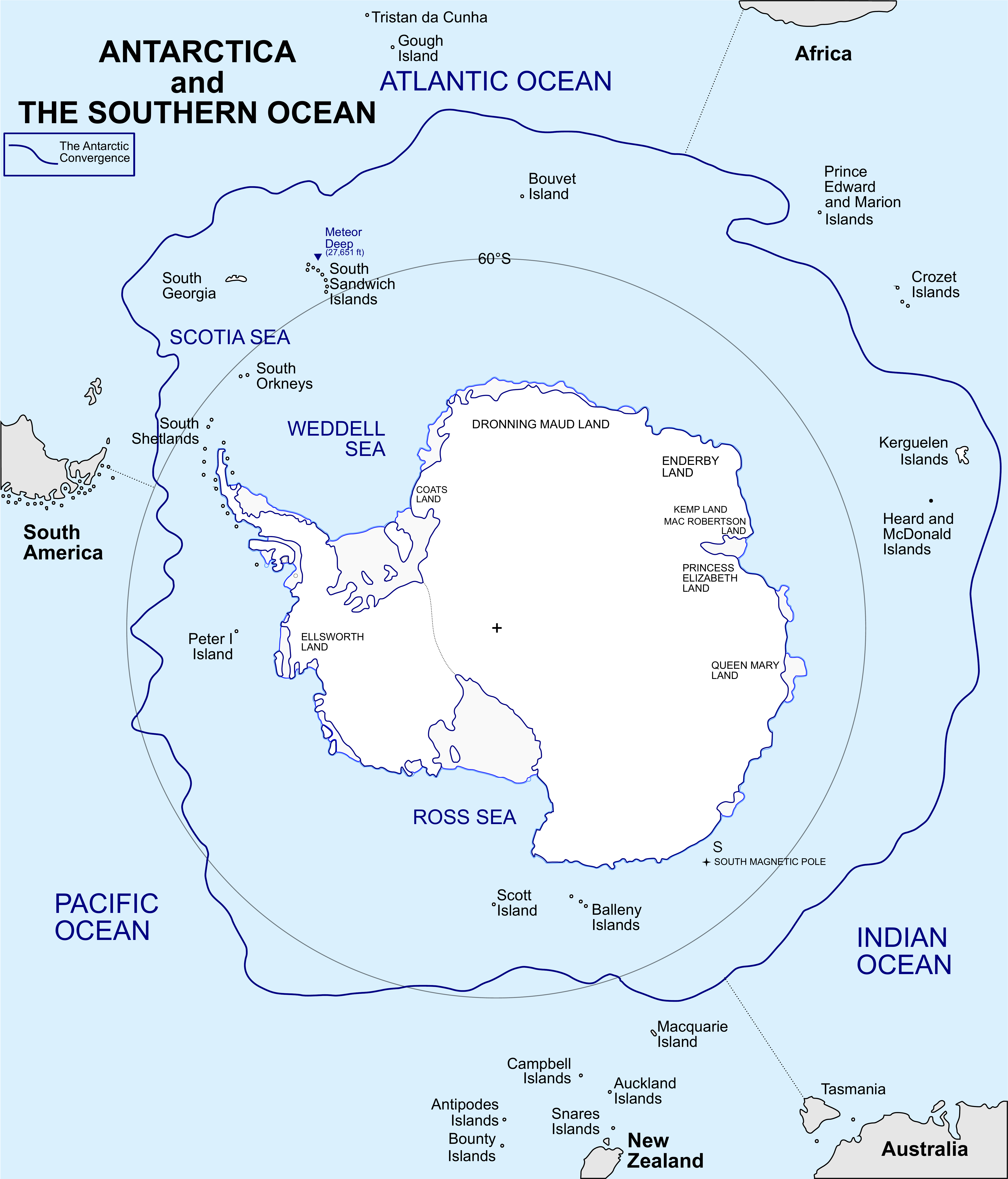}
    \caption{The Subantarctic Islands.}
    \label{fig:islands}
\end{figure}

Brown Skuas, Kelp Gulls, Southern Giant Petrels ({\it Macronectes giganteus}), and Snowy Sheathbills ({\it Chionis albus}) are all considered possible vectors of the virus in the subantarctic \cite{boulinier2023avian}.
Skuas are a predatory seabird that presented the first case of H5N1 in Antarctica \cite{Bennison01102024}.
NZ Skuas are migratory \cite{schultz2018non},
and they may be the main vector for HPAI in Antarctica \cite{bennett2024confirmation},
particularly as they kleptoparasitize other bird species and scavenge from carcasses.
HPAI affects skuas in particular – it has reduced the breeding population of Great Skua in Scotland by 66\% \cite{falchieri2022shift, camphuysen2022avian, tremlett2024uk}. Southern Giant Petrels and Wandering Albatrosses may also be vectors, as they are known to travel from SGSSI and PEI to NZ waters \cite{williams2006migrations}.

NZ’s bird populations are already vulnerable, having faced massive declines in bird populations, while a third of native bird species have become extinct since colonization \cite{wood2023post}. This study focuses on the subantarctic islands of NZ (SINZ), including 
Auckland Island,
Antipodes Island,
Bounty Island,
Campbell Island,
the Snares,
and also Macquarie Island, which is administered by Australia but in the same geographical region (see Fig.~\ref{fig:islands}\footnote{Adapted from https://commons.wikimedia.org/wiki/File:Antarctica\_and\_  
the\_Southern\_Ocean.svg CC-BY-SA 3.0 https://creativecommons.org/\\ 
licenses/by-sa/3.0/ © Hogweard. This image is also licensed CC-BY-SA 3.0.}). These islands have fragile ecosystems, and some taxa in the subantarctic islands have already gone extinct, such as the Auckland Islands Merganser (\textit{Mergus australis}) and the Macquarie Island Parakeet (\textit{Cyanoramphus novaezelandiae erythrotis}). HPAI presents a severe risk to their bird populations. 

Islands in the subantarctic are infrequently visited and understudied; the first comprehensive account of the birds of the Auckland Islands, for instance, was only published in 2019 \cite{lostgold}. Some bird populations have not been extensively surveyed, and estimates across the subantarctic may show wide variation (see Sec.~\ref{subsec:populations}), due largely to the cost of large expeditions in inhospitable regions. Community science datasets have already been appreciated as potential sources of information \cite{miskelly2020orn}. Below, we describe the datasets we used (Sec.~\ref{subsec:datasets}), the downloading (Sec.~\ref{subsec:downloading}) and filtering (Sec.~\ref{subsec:geofiltering}) process, our harmonization scheme (Sec.~\ref{subsec:harmonization}), and our statistical model (Sec.~\ref{subsec:model}). 

\begin{table*}[t]
\centering
\caption{Summary of datasets used in this study. * Auckland, Campbell, the Snares (** $\cap$ Subantarctic Is.), and Macquarie Islands are treated as geographic subsets of the Subantarctic Islands region.}

\label{tab:island-table}
\begin{tabular}{l*{12}{c}}
\hline \hline
   & \rotatebox{90}{All (Distinct)}
   & \rotatebox{90}{Crozets} 
   & \rotatebox{90}{Falkland Is.} 
   & \rotatebox{90}{SGSSI} 
   & \rotatebox{90}{Kerguelen I.} 
   & \rotatebox{90}{Marion \& PEI} 
   & \rotatebox{90}{Auckland Is.*} 
   & \rotatebox{90}{Campbell I.*} 
   & \rotatebox{90}{The Snares**} 
   & \rotatebox{90}{Macquarie Is.*} 
   & \rotatebox{90}{Subantarctic Is.} \\
\hline
Latitude   & -- & 
-46.33 & -51.80 & -54.42 & 
-49.24 & -46.90 & -50.74 & -52.55 & -48.02 & -54.63 & -54.50 \\
Longitude  & -- & 
51.55 & -59.30 & -36.56 & 
69.45 &  37.77 & 166.12 & 169.15 & 166.60 & 158.70 & 164.00 \\
Radius     &  -- &  
500  &   300  &   500  & 
250  &   500  &   250  &   250  &   250  &   250  &   800  \\ \\
\textbf{Observations} & & & & & & & & & & & & \\ \hline
eBird  & 319,652 &   1,132 & 182,225 &  68,393 &   5,644 &  11,226 &  15,668 &   9,075 &  10,558 &  13,580 &  51,032 \\
iNat  & 27,184 &    460 & 10,392 &  6,088 &  1,209 &    806 &  3,340 &  1,809 &  2,210 &  1,039 &  8,229 \\
GBIF  & 6,498 & 1,481 &   405 & 2,495 & 1,640 &    14 &    51 &    98 &    30 &   200 &   463 \\
RNBWS  & 2,171 &    19 &   603 & 1,121 &     9 &    73 &    80 &    64 &    80 &    70 &   346 \\
ASObs  & 215 &   0 &   0 &   0 &   0 &   0 &   0 &   0 &   0 &  61 & 215 \\
All Records  & 355,720 &   3,092 & 193,625 &  78,097 &   8,502 &  12,119 &  19,139 &  11,046 &  12,878 &  14,950 &  60,285 \\ \hline \hline
\end{tabular}
\end{table*}

\subsection{Community Datasets}
\label{subsec:datasets}

We used five distinct community science that include observational datasets across NZ and the Southern Oceans. These datasets vary widely in structure, taxonomy, temporal resolution, and observer methodology. A sixth, limited dataset of bird mortality numbers was sourced from the literature (Sec.~\ref{subsec:populations}). 

\textbf{1) eBird:} eBird is a global community science dataset maintained by the Cornell Lab of Ornithology \cite{sullivan2009ebird, SULLIVAN201431, ebirdHomepage}. Records include species names, observer effort metadata, spatial coordinates, timestamps, and standardized taxonomic labels based on Clements taxonomy \cite{clements2004}. Crucially, the ``complete'' protocol controls for effort, as observers indicate all species seen were reported, so each list also generates negative occurrence data for non-reported birds.

\textbf{2) iNaturalist:} A general-purpose biodiversity dataset with crowd-sourced species identifications \cite{iNatIDPage}. RG observations include verifiable images or audio, along with location, time, and taxa names. These names can be automatically suggested by ML trained on the dataset \cite{van2018inaturalist}, but are only ever assigned by human observers.

\textbf{4) Global Biodiversity Information Facility (GBIF):} GBIF is an aggregator, and contains \textgreater 3\,b records from \textgreater 110\,k datasets at present. 50\% of these records are the EOD, while 3.6\% are from iNaturalist.

\textbf{3) Royal Naval Birdwatching Society (RNBWS):} A dataset of 31\,k records consisting of maritime bird observations collected by British naval personnel \cite{rnbws}, with records generally from 1920-2024. These records include locations, species counts, time of day, and weather conditions.

\textbf{5) At-Sea Observations of Seabirds 1969 to 1990, Tasman Sea, New Zealand and Australian waters (ASObs):} A dataset of 48\,k at-sea observations of seabirds, dating from 1969--1990.
Seabirds were counted during 10-minute periods, like eBird's complete protocol. The data includes sighting locations, bird species, numbers and behavior, and an observation timestamp. 

\subsection{Downloading Process}
\label{subsec:downloading}

eBird's API \cite{ebirdAPI} does not allow large downloads, so datasets were downloaded using the eBird Request Data portal \cite{eBirdDownload}. Data were downloaded for Antarctica `AQ', Bouvet Island `BV', the Falklands `FK', SGSSI `GS', Heard and McDonald Islands `HM', New Zealand `NZ', Tasmania (Australia) `AU-TAS', Overseas France `TF', the High Seas `XX' (all pelagic regions over 200\,km from land \cite{eBirdXX}), and South Africa `ZA'. These regions were checked against land and sea polygons developed by \cite{schrimpf} at eBird  to filter data, and represent a balance between approximate ``closest point of land'' divisions for Economic Exclusive Zones or territorial waters and regions of higher turnover in avian communities (\cite{schrimpf}, pers. comm.). These polygons are currently available only to eBird reviewers, but the accessible regions listed above encompass all of the included areas.

iNaturalist has an API which can be used to search for relevant data \cite{iNatAPI}, but also has a native export functionality. We exported all RG bird observations from the study area with a search similar to \cite{iNaturalistSouthernAtlantic}. The queries are provided in Appendix 1. An equivalent mapping tool on GBIF \cite{gbifOccurrenceMap} was also used to select data for target areas to download. After more geographic filtering in R (see Sec.~\ref{subsec:geofiltering}), the derived dataset containing only the data which was used in our code was made available on GBIF  \cite{gbifDerivedDataset} using their derived dataset tool \cite{gbifDerivedCitationGuidelines}. We obtained a copy of the RNBWS database by requesting it from the society secretary \cite{rnbws}. The entire database for ASObs was downloaded as a Microsoft Excel spreadsheet from Te Papa's website \cite{ASObs}.


\subsection{Harmonization Pipeline}
\label{subsec:harmonization}

To integrate these disparate sources, we implemented a data harmonization workflow with the following stages:

\textbf{1) Download:} Each observation was downloaded and saved locally, as described in Section \ref{subsec:downloading}.

\textbf{2) Clean:} Some of the databases required extensive normalization and cleaning. In particular, the RNBWS and ASObs needed cleaning and standardization. The databases were initially composed of several Microsoft Excel files. For the RNBWS data, some of the spreadsheets had not yet been entered into the main database; these were edited to match the database format, and then included. For ASObs, the data was in several sheets, involving ship data and observation data. A Python \cite{Python} shim using pandas \cite{the_pandas_development_team_2024_13819579} merged rows together based on the record ID field. Then, the data went through the following process: 

\begin{itemize}
    \item Load dataset into OpenRefine
\cite{antonin_delpeuch_2025_15089184}.
    \item Normalize timestamps to UTC.
    \item Normalize scientific names:
    \begin{itemize}
        \item Remove unnecessary age, sex, and plumage information for ASObs.
        \item Correct common typographical errors in some names.
        \item Split subspecies IDs into a separate column.
        \item Use clustering tools and the integrated Wikidata reconciliation service to identify errors and dedupe names.
    \end{itemize}
    \item Repeat process for common names.
    \item Export to CSV format.
\end{itemize}

\textbf{3) Dataset filtering:} The GBIF database needed filtering, if not cleaning. All eBird and iNaturalist observations were removed as GBIF's dataset lags behind the available datasets from the respective platforms. All machine-created observations were also removed. These include observations like GPS tracking data for individual birds, and were considered too heterogeneous to include here.

\textbf{4) Geographic filtering:} All data was loaded back into R. Observations were reprojected into a uniform coordinate system (EPSG:4326), and then filtered to remove data outside of the study area. For more, see Sec.~\ref{subsec:geofiltering}, below.

\textbf{5) Taxonomic Reconciliation:} Some of the databases used different taxonomies for birds or identified individuals at different ranks. For instance, iNaturalist observations were often to the subspecies for Subantarctic Skua ({\it Stercorarius antarcticus lonnbergi}), a subspecies of Brown Skua. 

Only a few species of birds had known population and mortality data from HPAI, and there was a relatively low amount of species after geographic filtering. As a result, there was no expressed need for an automatic reconciliation between different taxonomies from eBird and iNaturalist. Instead, reconciliation was done by hand in R, although this would not be recommended for a larger dataset.

Some observations were made above the species level; for instance, eBird and iNaturalist both have an option for Southern/Northern Giant Petrel, where the genus but not the species was identified. eBird Best Practices \cite{ebirdBestPractices} suggests users drop these observations. As we only used complete checklists for eBird and not observation counts, they did not affect our model and were included. For iNaturalist and RNBWS, their inclusion may have affected the total observation counts. Subspecies identifications were ``rolled up'' (as eBird Best Practices suggests) to the species level during taxonomic reconciliation for all datasets. For Model 3 (Sec.~\ref{model3}) involving the Wandering Albatross, the entire species complex was rolled up into a single species to backfill sparse data, as older observations across datasets used a singular species identification for Wandering Albatross instead of one of the split species (see Sec.~\ref{subsubsection:target}).

\textbf{6) Taxonomic filtering:} Finally, we filtered for only those target species that we have mortality data for.

\subsection{Geographic filtering}
\label{subsec:geofiltering}

All data were imported into R \cite{rlang}, then filtered for circular regions with a radius of 200-500\,kms, drawn from central points for each island group in the dataset. The one exception was for SINZ including Macquarie Island (but excluding the Balleny Islands), where an 800\,km circle centered at 54\textdegree{}30$^\prime$00$^{\prime\prime}$S, 164\textdegree{}00$^\prime$00$^{\prime\prime}$E was used, as the islands were close enough together to overlap significantly. Circles for the Auckland, Campbell, and Macquarie Islands all fit within the SINZ circle. Part of the Snares circle covered Stewart Island / Rakiura, and was trimmed to the area that was an intersection of the subantarctic circle, to remove land-based observations. Centroid positions and diameter ranges are given in ~\ref{tab:island-table}.
Note that larger diameter circles did not necessarily lead to significantly more data, as pelagic data is often sparse.

Stewart Island / Rakiura was excluded from all datasets, as it could be considered part of the mainland due to its proximity to the South Island and its ecology. The St. Paul and Amsterdam Islands in the Indian Ocean, Tristan da Cunha and Gough Islands in the South Atlantic, Antipodes and Bounty Islands in NZ, and other islets were excluded due to their lower latitudes; Peter I Island off Antarctica was excluded due to its high latitude and distant location from the study area. The South Orkney Islands were excluded as being both underbirded and underrepresented in the literature, but close enough to SGSSI and the Falklands to make separate consideration unnecessary here.

Bouvet I., Heard and McDonald Is., and the Balleny Is. were excluded from the final analysis due to the lack of data. Across all datasets, there were only 211 observations from Bouvet, and only 419 from Heard and McDonald, and 1,388 records for the Balleny Is. 
Some birds do breed on the Balleny Is. 
\cite{francis2025circumpolar}, and may migrate north, 
but there are few records 
of vectors like 
skuas \cite{tidemann2015observations}.

\subsection{Population Metrics from the Literature}
\label{subsec:populations}

Accurate, recent population estimates for the subantarctic are rare. Here, estimates are given for individuals (not breeding pairs), along with mortality information regarding HPAI from two sources. Birds with minimal mortality counts were  not included. While observations from all birds were included in the dataset, only species listed below were used in the final model, as they were the only birds with known mortality counts.

\begin{table}[!t]
\caption{Summary of population statistics from SGSSI}
\label{table:sgssi}
\centering
\begin{tabular}{lrrrrrr}
\textbf{Species} & \textbf{Population} & \textbf{Mortalities} & \textbf{Mortality in \%}\\
\hline
Brown Skua     & 5,333 & 1,000 & 18.75 \\
Kelp Gull      & 4,000   & 100   & 2.50 \\
King Penguin   & 900,000 & 100   & 0.01 \\
Wandering Albatross  & 3,425 & 50 & 1.46 \\
\end{tabular}
\end{table}

\subsubsection{South Georgia and the South Sandwich Islands (SGSSI)}

Each of the mortality statistics came from a SCAR update \cite{scar2024}; earlier figures can be found in \cite{bennett2024confirmation}. Brown Skua population estimates of 2\,k breeding pairs were sourced from \cite{clarke2012important}, citing a 1983 count, and includes breeding adults and an added 1,333 birds for unpaired adults, extrapolating from \cite{Carneiro2014-qy}. Kelp Gulls and King Penguin counts includes only breeding pairs (here given as individuals, from the same count in \cite{clarke2012important}. Populations for Wandering Albatross were sourced from \cite{poncet2017recent}, and includes breeding pairs and chicks in order to align with mortality assessments from Marion, below. 

\subsubsection{PEI}

\begin{table}
\caption{Summary of population statistics from Marion and PEI. Mortality rates are only known for Marion.}
\label{table:marion}
\centering
\begin{tabular}{lrrrrrr}
\textbf{Species} & \textbf{Population} & \textbf{Mortalities} & \textbf{Mortality in \%}\\
\hline
Wandering Albatross & 10,184 & 150 & 1.47 \\
Brown Skua  & 1,225 & 80 & 6.53 \\
King Penguin & 134,000 & 120 &  0.09 \\
\end{tabular}
\end{table}

Around 3,000 pairs of Wandering Albatross breed on Marion \cite{vincent2008survival} (although that number is variable \cite{bonnevie2012effects}), and later estimates use 3,650 pairs across both Marion and Prince Edward Island \cite{ryan2009recent}. We use the latter number here, and estimate similar chick numbers across both islands. For Marion, with ``150 of approximately 1,900 [Wandering Albatross] chicks from the 2024 cohort having died'' \cite{marion2025web}, we assume the population has gone up to 3,800 pairs, and with a breeding success rate at Marion of 68\% \cite{Jones2017-yw}, the population of chicks can be calculated to be normally be 2,584 per year. As our data does not separate out the two islands, we double the amount of mortalities to 300 chicks.

Around 460 pairs of Brown Skua breed on the PEI Is. \cite{botw2020skua}. Skuas breed in pairs or trios, generally raise 2 chicks, and roughly a quarter of the non-breeding population are present on club sites during the breeding season \cite{Carneiro2014-qy}. Since adults were mentioned as casualties \cite{marion2025web}, an adult population of 1,225 will be used. A population of 67,000 pairs of King Penguin was estimated in 2009 \cite{Crawford01122009}. 

Southern and Northern Giant Petrels mortality numbers were estimated in \cite{marion2025web}, but the numbers were so low that these have been excluded from the dataset.

\subsubsection{Black-browed Albatrosses on the Falkland Islands}

On the Falkland Islands, there was a mass mortality event for 10,000 Black-browed Albatrosses ({\it Thalassarche melanophris}) \cite{scar2024}. The population is roughly 1.1\,m birds \cite{kuepfer2023influence}, which gives a mortality rate of 0.91\%. This species also breeds on SGSSI. A population decline at surveyed sites in 2014 gave a number of 18,298 pairs \cite{poncet2017recent}. Assuming an even rate of decline at non-surveyed sites, a population of 56,284 for all surveyed sites can be used as a proxy for this species on South Georgia. The Black-browed Albatross has around 42,000 breeding individuals on Campbell Island in SINZ \cite{nzbo2025bbal}.

\subsubsection{Target Species}
\label{subsubsection:target}

Given the known mortality metrics for bird species listed above, this model is only being applied for: Brown Skua, King Penguin, and Wandering albatross, based on SGSSI and Marion data; Kelp Gull, based on SGSSI data; and Black-browed Albatross, based on Falkland Is. data.

The Wandering Albatross is part of a species complex, containing Snowy Albatross (\textit{Diomedea exulans}) with Antipodean (\textit{D. antipodensis}), Gibson's Albatross (either \textit{D. antipodensis gibsoni} or \textit{D. gibsoni}), Tristan (\textit{D. dabbenena}), Southern Royal (\textit{D. epomophora}), and Northern Royal (\textit{D. sanfordi}). Here, Snowy Albatross data from SGSSI/Marion is applied to \textit{gibsoni} subspecies on the Auckland Islands and \textit{epomophora} on Campbell Island. The Black-browed albatross (\textit{Thalassarche melanophris}) data is applied exclusively to the NZ subspecies \textit{T. m. impavida}, which is treated as a full species by some authorities \textit{T. impavida} \cite{osnzcc2022}, and which breeds on Campbell Island. Kelp Gull data from SGSSI was applied solely to the subantarctic islands of NZ. King Penguin data was applied only to the Crozets, Kerguelen, and Macquarie Island.

Another possible vector for HPAI from Antarctica, Snowy Sheathbills \cite{boulinier2023avian}, are not considered here, as sheathbills are not found in SINZ.

\subsection{Applying the model}
\label{subsec:model}

To estimate the population size of a species in an area where no or few population surveys exist, like SINZ (Area~B), we used a calibration-based extrapolation from a reference area, like SGSSI (Area~A), where observational data, population estimates, and mortality rates are available.

Let
\begin{itemize}
  \item $O_{s,A}^{(i)}$ be the number of observations of species $s$ in Area~A from dataset $D_i$ (within the associated geographic polygon in any source platform $S$),
  \item $\sum O_A^{(i)}$ be the total number of bird observations in Area~A from dataset $D_i$ for $S = \left\{ \text{iNaturalist}, \text{GBIF}, \text{RNBWS} \right\}$,
  \item $\sum C_{A}^{(i)}$ be the total amount of checklists in Area~A from dataset $D_i$ for $S = \left\{ \text{eBird},  \text{ASObs} \right\}$,
  \item $P_{s,A}$ be the known population of species $s$ in Area~A,
  \item $O_{s,B}^{(i)}$, $\sum C_{B}^{(i)}$ and $\sum O_B^{(i)}$ be the same, for Area~B,
  \item $M_{s,A}$ be the number of observed mortalities of species $s$ in Area~A,
  \item $w_i$ be the weight assigned to dataset $D_i$.
\end{itemize}

We first compute the relative observation frequency for each dataset ($S = \left\{ \text{eBird}, \text{ASObs} \right\}$):
\[
f_A^{(i)} = \frac{O_{s,A}^{(i)}}{\sum C_A^{(i)}}, \quad f_B^{(i)} = \frac{O_{s,B}^{(i)}}{\sum C_B^{(i)}}
\]

For $S = \left\{ \text{iNaturalist}, \text{GBIF}, \text{RNBWS} \right\}$, where effort is not controlled, we instead use:

\[
f_A^{(i)} = \frac{O_{s,A}^{(i)}}{\sum O_A^{(i)}}, \quad f_B^{(i)} = \frac{O_{s,B}^{(i)}}{\sum O_B^{(i)}}
\]

Because SGSSI and Marion differed in their observed mortality rates over time, we also combined data from both areas and applied a weighting scheme (in all cases, .8 for SGSSI, and .2 for Marion):

\[
f_A^{(\text{weighted})} = \frac{\sum_i w_i \cdot f_A^{(i)}}{\sum_i w_i}, \quad f_B^{(\text{weighted})} = \frac{\sum_i w_i \cdot f_B^{(i)}}{\sum_i w_i}
\]

Assuming constant detectability per individual and species between areas and across datasets, we estimate the unknown population size in Area~B as:
\[
\hat{P}_{s,B} = \frac{f_B^{(\text{weighted})}}{f_A^{(\text{weighted})}} \cdot P_{s,A}
\]

When we had numbers for the mortality rate in Area~A, we extended the population estimate to infer possible mortality in Area~B, under similar conditions. Where population numbers for a target Area~B were known, they were substituted for $\hat{P}_{s,B}$.

We combined datasets in three separate models. Model 1 used only eBird Data, as it is both the largest and the most quality-controlled dataset in our study. Model 2 integrated the second largest dataset, iNaturalist, together with Model 1. Model 3 combined RNBWS, ASObs, and GBIF data with both datasets in Model 2 for a subset of the species given above, depending on the amount of data and the disparities between $f^{(i)}$.






\section{Results}
\label{results}

\subsection{Model 1: eBird}
\label{model1}

The first model uses only eBird data.

\textbf{Brown Skua}: Without weights for SGSSI and Marion, the predicted population for SINZ Brown Skuas is 2,761 individuals, with a mortality count of 429 (15.53\%). Assigning a .8 weight to SGSSI data,  the population size increases to 3,545, with a mortality count of 629 (17.75\%). Assuming similar conditions to SGSSI, this is a higher extrapolated population size than the 500-1,000 pairs suggested in \cite{nzbirdsonlineskua}. Assuming 750 breeding pairs and a total population size of 2\,k birds, a weighted model results in a mortality count of 326 (16.31\%). \textbf{King Penguin}: Without weights, for King Penguins on Macquarie, the extrapolated population size of 612\,k. Known populations on Macquarie point to closer to 70\,k breeding pairs \cite{martinez2020king}, although this may be decreasing  and counts of only breeding pairs obscures population dynamics \cite{pascoe2022current}. Using this number as a base, a weighted model results in a mortality count of 38 (0.03\%). A weighted model for 300\,k breeding pairs on the Crozet Is. and 260\,k breeding pairs on Kerguelen \cite{martinez2020king} results in counts of 161 and 139, respectively. 
\textbf{Wandering Albatross}: A weighted model based on Snowy Albatross on SGSSI and Marion for a known population of 6,550 Antipodean Albatross on Auckland Island predicts a mortality count of 96 (1.46\%). The same model for 15.6\,k Southern Royals on Campbell Island predicts a mortality count of 228 (1.46\%). \textbf{Black-browed Albatross}: Using only eBird data and a known population of 42\,k birds on Campbell I., we predict a mortality count of 382 birds (0.91\%). \textbf{Kelp Gull}: For Macquarie, with a known population of ~150 birds \cite{micf2025kelp}, we predict a mortality count of 4 birds. Small populations of Kelp Gull breed on the subantarctic islands \cite{miskelly2022birds}. A weighted model predicts a total population of 3,133 birds for SINZ, with a mortality count of 78 (2.5\%).


\subsection{Model 2: eBird and iNaturalist}
\label{model2}

The second model combined eBird and iNaturalist data. For known populations, the predicted mortality rates were identical to Model 1.

\textbf{Brown Skua}: A weighted combined model with .8 for SGSSI and .2 for Marion data predicted a  slightly higher population of 3,565 skuas, with a mortality rate of 632 (17.74\%). \textbf{King Penguin}: A similarly weighted model with no known populations predicted  650\,k birds on Macquarie, but a closer numbers of 781\,k on the Crozets and 555\,k on Kerguelen, with mortalities of 295 and 210, respectively. The \textbf{Wandering Albatross} and \textbf{Black-browed Albatross} extrapolated populations were either too low -- 1.7\,k for Antipodean, 7.8\,k for Southern Royal Albatross -- or too high -- 1.4\,m for Black-browed Albatross on Campbell I. \textbf{Kelp Gull} was very similar to Model 1: 3,106 individuals in SINZ, with a mortality of 78 birds. 


\subsection{Model 3: All datasets}
\label{model3}

For RNBWS, the observation counts and ratios $\sum O_A^{(i)}$ were 
off from eBird and iNaturalist by orders of magnitude. This is almost certainly due to the sparse data in this dataset, as shown in Tab.~\ref{tab:island-table}. There was one exception: Wandering Albatross (67 out of 1,121 observations, 5.98\%) from SGSSI, compared to 15.38\% from eBird. This observation included five observations of Snowy, Southern Royal, and generic Royal Albatrosses. Likewise, for GBIF data, the counts and ratios were often off by at least half an order of magnitude. We used only Wandering Albatross totals from SGSSI (113/2495, 4.53\%). The observation ratios from GBIF to eBird for Black-browed (12.53\% to 40.49\%) and Snowy Albatrosses (2.16\% to 1.61\%) were similar for SINZ, but not for the smaller polygons for the Campbell or Auckland Is. The ASObs database only includes information from waters near Australia and NZ. For all three extra databases, the data on Kelp Gull, Brown Skua, and King Penguin was too sparse to be considered in the final runs.

\textbf{Wandering Albatross}: Using only using only data from eBird, iNat, RNBWS, and GBIF from SGSSI, and applying with observation ratio frequencies from eBird, iNat, ASObs, and GBIF in SINZ, we predict a final population of 7,507, with a mortality count of 110 and a mortality rate of 1.46\%. \textbf{Black-browed Albatross}: Using the same databases and sourcing ratios from the Falkland Is., a far-too-high extrapolated population of 961\,k was predicted, with a mortality rate of 8,738 (0.91\%). 

\section{Discussion}
\label{discussion}

Some of the projections from the three models explored above are novel and interesting. Where the numbers were similar between Model 1 and 2, the latter was stronger, as it came from multiple datasets. The population of Skuas appears stable at around 3,550, which is only slightly higher than the projected 1-2\,k estimate. There is no exact data on population sizes for Brown Skuas in NZ: ``no recent or overall data available, but likely 500-1,000 breeding pairs, trios or groups in the NZ region. The Chatham, Snares and Stewart Island populations often breed in trios or larger groups, but co-operative breeding is rarely reported from the more southerly subantarctic islands.'' \cite{nzbirdsonlineskua} The projected population is roughly the expected size, given that trios and unpaired birds may suggest a higher population than in the quoted estimate. 

For known populations of birds, the proportional mortality rates are uncontroversial. The weighting -- .8 for SGSSI, and .2 for Marion -- allows for spurious data to be normalized, but does not resolve the issue that the difference in mortality rates largely reflects time of exposure. Marion's first case was at the beginning of the austral summer of 2024/25, while SGSSI's was almost a year earlier \cite{scar2024, marion2025web}. Assuming similar circumstances within genera and across oceans, one could expect that HPAI could affect birds as much in the subantarctic as in Britain. For skuas, the extrapolation would be a potential mortality rate of 66\%, or 2,353 of the population of 3,565 skuas projected by Model~2 in SINZ. The story is incomplete with only two datapoints in the subantarctic of SGSSI and Marion, or only one data point of the Falklands.

Brown Skuas are currently assessed as Least Concern globally by the IUCN \cite{birdlife2018catharacta}. 
However, the NZ population is ``Nationally Vulnerable'' according to the New Zealand Threat Classification System, which notes that it has already declined and that the current population is comprised of 2\,k birds across the entire country (and not just in the subantarctic islands) \cite{robertson2021conservation}. The rating was already downgraded in 2021 from the last assessment in 2016 \cite{robertson2016conservation} when they were classified as ``Nationally Uncommon'', based on a reinterpretation of the data, a slow generation time of 12 years \cite{bird2020generation}, and their interactions with humans.  A population size of 1,212 birds may lead to local extinctions at some of the subantarctic islands, and may necessitate a new classification as ``Nationally Endangered'' or ''Nationally Critical'', the two most severe classifications before ``Extinct''. The subspecies in NZ, {\it Stercorarius antarcticus lonnbergi}, also breeds in Antarctica; the two other subspecies breed on Tristan da Cunha and Gough Island, and in Argentina, respectively \cite{botw2020skua}. However, local extirpations on an island may be irreversible without extensive human effort.

Our population metric of 3,565 is an informed one, assuming similar conditions of observation and distribution of birds between SINZ and SGSSI and PEI. This is not necessarily a given; but it is more demographic information than is currently available through other sources. Any model depends upon the strength of the data put into it. In this case, the assumption that some islands have similar conditions to others resulted in vast over-prediction for Black-browed Albatrosses, where the breeding population on the Falklands is much higher than the breeding population in SINZ, which occurs only on Campbell Island \cite{nzbirdsonlinebb}. Likewise, King Penguin colonies, and possibly Wandering Albatross, are too variable for this extrapolation.

However, the Kelp Gull extrapolated populations appear as stable as Skuas, and may be closer to accurate for the region. The Kelp Gull is widely distributed across the southern hemisphere. The subspecies in NZ is unclear, and may be an endemic subspecies \textit{Larus dominicanus antipodum} \cite{osnzcc2022, Littauer2025NotornisKaroro}.
Both the IUCN and the NZTCS consider it Not Threatened \cite{birdlife2018larus, robertson2021conservation} -- the NZTCS assessment is only for birds in NZ. Neither of these assessments apply to Kelp Gull populations in the subantarctic islands, which may have vagrant gull populations and which may repopulate following population collapse. 

These numbers can be improved upon. In particular, more fine-tuned knowledge could be applied to each species and distribution, and temporal variation for HPAI infection and more thorough filtering of data could also lead to different results. The use of the RNBWS, GBIF, and ASObs databases showcases how small datasets applied indiscriminately may magnify issues in the underlying data. Both ASObs and RNBWS were at-sea datasets, and the distribution of birds differs at sea versus at an island colony. Pelagic datasets may need different treatment from terrestrial ones. The ASObs 10-minute protocol was functionally identical to eBird complete checklists, and if the restrictive data attribution license was removed, it could easily be added into eBird. GBIF already aggregates all of the data together; but this study has shown clearly that while GBIF allows for use of many datasets, for observational data eBird and iNaturalist dominate. When these datasets were removed from GBIF in preprocessing, Model 3 still suffered in comparison with the previous two. As such, unrealistic population estimates can serve as useful indicators to identify issues in the underlying community data.

There are other biases. In section \ref{sec:community-science}, we mentioned some of the biases inherent in community science datasets. There are further biases to this particular model:
\begin{itemize}
    \item Inferences from sparse data are not robust. As shown in Table \ref{tab:island-table}, the datasets are not large for any of these islands. These statistics are also combined for all species, making accurate inferences less reliable.
    \item Platform differences impact data. iNaturalist observations skew towards charismatic species that are easy to photograph: placid albatross chicks or penguins are more likely to be included than shy species like snipe, even on islands with birds unused to human populations.
    \item Each island is different, and has its own set of geophysical and ecological properties which should be accounted for in a more rigorous model.
    \item Species differences. Mortality rates for chicks of an albatross impact populations differently than for skuas. For instance, albatrosses breed in alternative years, and skuas can breed in trios.
    \item Seabirds can die at sea. Shore counts are rarely systematic on isolated islands, and neither study presented protocols for mortality counts \cite{scar2024, marion2025web}.
    \item HPAI may not affect each population in the same way.
\end{itemize}

The known population numbers used in this paper are best effort estimates, as surveying penguins, seabirds, and nesting birds is difficult to  due to the bird phenology, and simple count surveys should only be used over long periods of time \cite{frederick2006estimating, foley2018king}. Community science platforms like eBird provide count protocols for observers, but these protocols are not as stringent as those used by professional observers \cite{robinson2021benchmark, feng2021comparing, matthiopoulos2022integrated}, and the resulting data may have wider margins for error.

The model shown here showcases how multiple datasets can be used to indicate population numbers for unknown areas, using community science data. It also shows how those numbers can inform epidemiological studies of bird populations. Further, the workflow explored in Sec. \ref{subsec:harmonization} can be applied elsewhere. Epidemiology of wild bird populations is not the only possible case study for this work. For instance, climate change adds another significant stressor to birds (\cite{weinhaupl2024potential}; but see \cite{BourkeScottD2024Pioc}). This is also the case for Antarctica (\cite{croxall2002environmental}; \cite{sauser2021sea}; \cite{mcquaid2023understanding}; etc.). These threats, along with introduced predators and extractive industrial practices must be considered to fully appreciate the compounding effect of HPAI, climate change and habitat loss on bird resilience. This model does not yet account for such potential uses of community science data.

\section*{Acknowledgment}
Ethics: No ethics waivers were needed for this work.
Funding: This work was unfunded.
Data: ASObs data is made available by the Museum of New Zealand
Te Papa Tongarewa (Te Papa) under a CC-BY 4.0 Int. license \cite{tepapa}.
Code: All related code has been made available at https://codeberg.org/RichardLitt/subantarctic-birds/.
AI: ChatGPT was consulted as a guide during code production for this paper, and to scaffold the formatting for Tab.~\ref{tab:island-table} and the LaTeX equations in Sec.~\ref{subsec:model}. We are fully responsible for all content in this paper, and any mistakes are our own.


\bibliographystyle{IEEEtran}
\bibliography{citations, gbif-data}


\end{document}